\documentstyle[12pt]{article}
\advance \textheight by \topskip
\makeatletter
\def\eqnarray{\stepcounter{equation}\let\@currentlabel=\theequation
\global\@eqnswtrue
\global\@eqcnt\z@\tabskip\@centering\let\\=\@eqncr
$$\halign to \displaywidth\bgroup\@eqnsel\hskip\@centering
  $\displaystyle\tabskip\z@{##}$&\global\@eqcnt\@ne 
  \hfil$\displaystyle{\hbox{}##\hbox{}}$\hfil
  &\global\@eqcnt\tw@ $\displaystyle\tabskip\z@
  {##}$\hfil\tabskip\@centering&\llap{##}\tabskip\z@\cr}
\@addtoreset{equation}{section}
  \def\theequation{\thesection.\arabic{equation}}
\makeatother
\mathchardef\by="0202

\begin{document}

\thispagestyle{empty}
\begin{flushright}
%{\bf DFTUZ/95/26}\\
%{\bf DFUPG-106/95}\\
{\bf hep-th/9511072}
\end{flushright}
$\ $\vskip 2truecm
\begin{center}

{ \Large \bf A Pseudoclassical Model for $P,T-$Invariant}\\
\vskip0.5cm
{\Large \bf Planar Fermions}\\
\vskip1.0cm
{ \bf G. Grignani${}^{a,}${}\footnote{E-mail: 
grignani@perugia.infn.it},
M. Plyushchay${}^{b,}${}\footnote{On leave from the
{\it Institute for High Energy Physics, Protvino,
Moscow Region, Russia}; e-mail: 
mikhail@posta.unizar.es}
and 
P. Sodano${}^{a,}${}\footnote{E-mail: sodano@perugia.infn.it}\\[0.3cm]
{\it ${}^{a}$Dipartimento di Fisica and Sezione I.N.F.N.}\\
{\it Universit\'a di Perugia, via A. Pascoli, I-06100 Perugia, Italy}}\\
{\it ${}^{b}$Departamento de F\'{i}sica Te\'orica, Facultad de Ciencias}\\
{\it Universidad de Zaragoza, 50009 Zaragoza, Spain}\\ 
\end{center}
\vskip2.0cm
\begin{center}                            
{\bf Abstract}
\end{center}
A pseudoclassical model is proposed for the description of
planar $P,T-$invariant massive fermions.  The quantization of
the model leads to the (2+1)-dimensional $P,T-$invariant fermion
model used recently in $P,T-$conserving theories of high-T${}_c$
superconductors.  The rich symmetry of the quantum model is
elucidated through the analysis of the canonical structure of
its pseudoclassical counterpart. We show that both the quantum
$P,T-$invariant planar massive fermion model and the proposed
pseudoclassical model --- for a particular choice of the
parameter appearing in the Lagrangian --- have a U(1,1)
dynamical symmetry as well as an $N=3$ supersymmetry.  The
hidden supersymmetry leads to a non-standard superextension of
the (2+1)-dimensional Poincar\'e group.  In the quantum theory
the one particle states provide an irreducible representation of
the extended supergroup labelled by the zero eigenvalue of the
superspin.  We discuss the gauge modification of the
pseudoclassical model and compare our results with those
obtained from the standard pseudoclassical model for massive
planar fermions.
\begin{center}
{\bf To appear in Nucl. Phys. B}
\end{center}
\newpage

\section{Introduction}

Planar gauge field theories have many interesting 
features \cite{i1}. Especially in their use for constructing
models of high-T${}_c$ superconductors \cite{8}
one is interested in having a 
$P-$ and $T-$invariant system of topologically massive gauge fields
and a $P-$ and $T-$invariant system of massive Dirac spinor fields.
For this  purpose, one usually introduces doublets of these fields with 
mass terms having opposite sign \cite{8,i8}.
In a recent paper \cite{7} the hidden dynamical symmetries
of the simplest $P-$ and $T-$invariant free fermion theory were 
investigated. The model \cite{7} is described by the Lagrangian 
\begin{equation}
{\cal L}=\bar{\psi}_u(p\gamma+m)\psi_u
+\bar{\psi}_d(p\gamma-m)\psi_d
\label{i1}
\end{equation}
and it is invariant under a global U${}_c$(1)
symmetry describing chiral rotations to which it is 
associated the conserved chiral current $I_\mu=
\frac{1}{2}(\bar{\psi}_u\gamma_\mu\psi_u-\bar{\psi}_d\gamma_\mu\psi_d).$
A very interesting feature of the model described by
(\ref{i1}) is the appearance of a hidden $N=3$ supersymmetry
which yields a non-standard superextension of the Poincar\'e 
group so that the one particle states associated to
(\ref{i1}) realize an irreducible representation 
of the Poincar\'e supergroup, labelled by the zero eigenvalue
of the superspin operator \cite{7}.

In a beautiful paper \cite{1}  
Gibbons, Rietijk and van Holten investigated 
space-time  symmetries in terms of the 
motion of pseudoclassical spinning point 
particles \cite{2,3,4}. In their analysis they
revealed the existence of a non-standard
supersymmetry characterized by the fact that the
Poisson brackets of the odd Grassmann generators yield
an even integral of motion different from the Hamiltonian 
of the system.
A non-standard supersymmetry of the same kind is
useful to describe the hidden symmetries \cite{5} of
a 3-dimensional monopole \cite{6}.
  
In this paper we shall show that the hidden $N=3$ supersymmetry
of the $P,T-$invariant 3D fermion model described by 
(\ref{i1}) may be understood in a natural way 
starting with the pseudoclassical model
of a relativistic spinning particle proposed in 
ref. \cite{9}. 

The pseudoclassical model of ref. \cite{9}
is defined by the Lagrangian
\begin{equation}
L=\frac{1}{2e}(\dot{x}_\mu-iv\epsilon_{\mu\nu\lambda}
\xi^{\nu}\xi^{\lambda})^2 -\frac{e}{2}m^2 -
2\nu m v \theta^+\theta^{-}-\frac{i}{2}\xi^{\mu}\dot{\xi}_\mu
+\frac{i}{2}\theta_a\dot{\theta}_a.
\label{i2}
\end{equation}
A complete description of the quantities appearing
in (\ref{i2}) is given in section 2.
Here we only notice that the pseudoclassical model
(\ref{i2}) depends on a parameter $\nu$.
As we shall see,  there is a special value of the parameter
$\nu$ ($\nu=1$) which leads to a maximal symmetry of the
classical system.
 
Since the pseudoclassical model (\ref{i2})
has an even nilpotent constraint, it is similar
to the class of pseudoclassical models used in describing 
particles  with spin $s\geq 1$ \cite{10}.
Its main difference with the standard 
pseudoclassical model (SPM)
for the  massive Dirac particle \cite{3}
lies in the fact that the odd constraint of the latter model
is replaced in \cite{9} by the even nilpotent constraint.
We shall show that, although at the quantum level
both models lead to the $P,T-$invariant planar model
described by (\ref{i1}), only the model of ref. \cite{9}
reproduces classically the same symmetries of its quantum
counterpart.
{}For the 3D standard pseudoclassical model \cite{3} the
situation is quite reminiscent of what occurs in systems with a
quantum anomaly, since the classical symmetries are not
preserved by the quantization procedure.  Unlike the SPM
\cite{3}, the model considered in this paper naturally yields a
U(1) gauge theory with the gauge field coupled to the U${}_c$(1)
chiral current and it allows for a nontrivial interplay between
local and discrete symmetries on one hand, and ordinary and
graded symmetries on the other hand.
   
The paper is organized as follows.
In order to have a self-contained presentation, we introduce in
section 2 the pseudoclassical model of 
ref. \cite{9}. In this section we analyze its
global, local and discrete symmetries within
the framework of the Lagrangian formalism.
Section 3 is devoted to the Hamiltonian description of the
system. 
First we get the equations of motion 
and find the integrals of motion which are the generators 
of the continuous symmetries described in section 2.  Then, we 
notice that for a
special value of the parameter of the 
pseudoclassical model ($\nu=1$)
the system has two additional (local in the evolution parameter)
integrals of motion.
We find that the integrals
of motion of the system form a broader U(1,1) symmetry and
a hidden $N=3$ supersymmetry, and present both symmetries 
in a Poincar\'e covariant form. 
Section 4 is devoted to gauging 
the U(1) symmetry corresponding at the quantum level to the chiral
rotations.
In section 5 we quantize the model.
We show that $\nu=1$ is also 
special from the point of view of the quantum theory:
namely, we demonstrate that only for $\nu=1$
the quantization procedure preserves the $P-$ and $T-$ symmetries 
of the classical model.
For this value of the parameter $\nu$ 
the 
U(1,1) symmetry and the $N=3$ supersymmetry are realized
in the quantum theory 
and the $N=3$ supersymmetry leads to the non-standard
superextension of the Poincar\'e group.
Finally, we investigate the quantum theory of the model
with chiral U(1) gauge symmetry.
Section 6 is devoted to some concluding remarks.

In Appendix A we evidence a hidden symmetry
of the pseudoclassical model existing
in the subspace of the Grassmann
(pseudo)scalar variables. 

The analysis of the planar SPM \cite{3}
and its dynamical symmetries
is presented in Appendix~B.

\section{The model and its Lagrangian symmetries}
\subsection{The model}

The planar model analyzed in ref. \cite{9} is
defined by the action \begin{equation}
A=\int_{\tau_i}^{\tau_f}L d\tau +B,
\label{2}
\end{equation}
where $L=L_0+L_\xi+L_\theta$ is the Lagrangian,
\begin{equation}
L_0=\frac{1}{2e}(\dot{x}_\mu-iv\epsilon_{\mu\nu\lambda}
\xi^{\nu}\xi^{\lambda})^2 -\frac{e}{2}m^2 -
2\nu m v {\cal N},
\label{3}
\end{equation}
\begin{equation}
{\cal N}=\theta^+\theta^-=-i\theta_1\theta_2,
\label{4}
\end{equation}
\begin{equation}
L_\xi=-\frac{i}{2}\xi_\mu\dot{\xi}{}^{\mu},\qquad
L_\theta=\frac{i}{2}\theta_a\dot{\theta}_a,
\label{5}
\end{equation}
and
$B=B_\xi+B_\theta$ is the boundary term,
\begin{equation}
B_\xi=-\frac{i}{2}\xi_\mu(\tau_f)\xi^\mu(\tau_i),\qquad
B_\theta=\frac{i}{2}\theta_a(\tau_f)\theta_a(\tau_i).
\label{6}
\end{equation}
In (\ref{3}) $m$ is a mass parameter and
$\nu\neq 0$ is a dimensionless
parameter. We use the metric $\eta_{\mu\nu}=diag(-,+,+)$ and the
totally antisymmetric tensor $\epsilon_{\mu\nu\lambda}$,
$\epsilon^{012}=1$.

The configuration superspace of the
model is described by the set of
variables $x_\mu$, $\xi_\mu$, $\theta^\pm$, $e$ and
$v$.
We denote with $x_\mu$ the space-time coordinates of the
particle; together with the scalar Lagrange multipliers
$e$ and $v$ they are real even variables.
$\xi_\mu$ is a real odd
(Grassmann) vector whereas $\theta^+$ and $\theta^-$ are
mutually conjugate odd scalar variables 
related to the real variables $\theta_a$, $a=1,2,$
by the equation
\begin{equation}
\theta^\pm=\frac{1}{\sqrt{2}}(\theta_1\pm
i\theta_2).
\label{1}
\end{equation}
We shall specify later the transformation properties of all the variables
under $P$ and $T$ inversions.

The form of the kinetic terms $L_\xi$ and $L_\theta$ (as well as
that of the boundary terms $B_\xi$ and $B_\theta$) 
manifests the fact 
that the scalar
variables $\theta_a$, $a=1,2$, are ``timelike'' in contrast to
the one ``spacelike'' scalar variable $\xi_*$ used in the standard
pseudoclassical formulation of the massive spin-1/2 particle
(see Appendix B).

The inclusion of the boundary terms
in the action (\ref{2}) is needed since
the equations for the 
Grassmann variables are first order \cite{11}. As a result,
the classical solutions to these equations 
extremize the action (\ref{2}) with the boundary
conditions:
\begin{equation}
\delta\eta_A(\tau_f)+\delta\eta_A(\tau_i)=0,\quad
\eta_A=\xi_\mu,\theta_a.
\label{7}
\end{equation}

The change of variables 
$
\theta_1,\,\theta_2\rightarrow \theta_1,\,-\theta_2,
$
and, correspondingly, $\theta^\pm\rightarrow \theta^\mp$,
amounts only
to changing ${\cal N}\rightarrow -{\cal N}$ in the 
action (\ref{2}). 
Without loss of generality, one may then always
choose
\begin{equation}
\nu>0,
\label{8}
\end{equation}
since the model with $\nu<0$ is reproduced by the above
change of the variables.

\subsection{Global symmetries}

The action (\ref{2})
is Poincar\'e-invariant. In addition, it is invariant
under the following global super-transformations:
\begin{equation}
\delta\xi_\mu=\epsilon_\mu,
\qquad
\delta x_\mu=2i\epsilon_{\mu\nu\lambda}\epsilon^{\nu}\cdot
\int_{\tau_i}^{\tau} v(\tau')\xi^{\lambda}(\tau')d\tau',
\label{9}
\end{equation}
where $\epsilon_\mu$ is a constant odd vector infinitesimal
transformation parameter.
Unlike the ordinary global super-transformations,
the transformations (\ref{9}) have a nonlocal character
in the evolution parameter.  As a consequence,
\[
(\delta_1\delta_2-\delta_2\delta_1)x_\mu=
-4i\epsilon_{\mu\nu\lambda}\epsilon^\nu_1\epsilon^\lambda_2
\cdot\int_{\tau_i}^{\tau}v(\tau')d\tau'\neq const.
\]
Thus, the commutator of two global
super-transformations (\ref{9}) does not yield the
space-time translations characterized by a constant nilpotent
vector as it happens for the standard global
super-transformation in systems  with
a supersymmetric spectrum.

The action is invariant
under another set of global super-transformations
\begin{equation}
\delta_\epsilon x_\mu=i\epsilon\xi_\mu,\qquad
\delta_\epsilon\xi_\mu=\epsilon e^{-1}(\dot{x}_\mu
-iv\epsilon_{\mu\nu\lambda}\xi^{\nu}\xi^{\lambda})
\label{10}
\end{equation}
with a constant scalar odd parameter $\epsilon$.
In the standard
pseudoclassical model of a spin-1/2 particle \cite{3},
there is 
{\it local} supersymmetry (with a local infinitesimal
parameter $\epsilon=\epsilon(\tau)$)
induced by 
transformations similar in form to eq. (\ref{10}).

The action is also invariant under
\begin{equation} 
\delta_\epsilon
\xi_\mu=i\epsilon
\epsilon_{\mu\nu\lambda}\xi^{\nu}\xi^{\lambda}
\label{11}
\end{equation}
with the odd scalar infinitesimal transformation
parameter $\epsilon$.

Furthermore, there is 
invariance under the
global U(1) transformations
\begin{equation}
\delta_\gamma\theta^{\pm}=\pm i\gamma \theta^{\pm}
\label{12}
\end{equation}
(or corresponding SO(2) transformations in terms of the real
variables $\theta_a$),
and the global even transformations
\begin{equation}
\delta_{\omega} x_\mu
=i{\omega}\epsilon_{\mu\nu\lambda}\xi^\nu\xi^\lambda,
\qquad
\delta_{\omega}\xi_\mu=2{\omega} e^{-1}
\epsilon_{\mu\nu\lambda}\dot{x}{}^{\nu}\xi^{\lambda},
\label{13}
\end{equation}
characterized by
the even scalar transformation parameters
$\gamma$ and ${\omega}$, respectively.
In Section 3 we shall analyse these global symmetries
within the Hamiltonian formalism.

\subsection{Local symmetries}

There are two local symmetries of the action.
They are given by
the transformation
\begin{eqnarray}
&\delta_\alpha E=\frac{d}{d\tau}(\alpha E),\qquad
E=e,v,&\nonumber\\
&\delta_\alpha X=\alpha\dot{X},\qquad
X=x_\mu,\xi_\mu,\theta_a,&
\label{14}
\end{eqnarray}
which is the reparametrization transformation,
and by the transformation
\begin{eqnarray}
&\delta_\beta e=0,\qquad
\delta_\beta v=\dot{\beta},\qquad
\delta_\beta\theta^{\pm}=\pm im\beta\theta^{\pm},&
\nonumber\\
&\delta_\beta x_\mu=i\beta\epsilon_{\mu\nu\lambda}
\xi^{\nu}\xi^{\lambda},\qquad
\delta_\beta\xi_\mu=2\beta e^{-1}\epsilon_{\mu\nu\lambda}
\dot{x}{}^{\nu}\xi^{\lambda},&
\label{15}
\end{eqnarray}
to which we shall refer to as the `$\beta$-transformation'.
In eqs. (\ref{14}), (\ref{15}) 
$\alpha$ and $\beta$ are the infinitesimal transformation
parameters. Since we have $\delta_\alpha L=(d/d\tau)(\alpha L)$
and
$\delta_\beta L=(d/d\tau) (i\beta
e^{-1}\epsilon_{\mu\nu\lambda}\dot{x}{}^{\mu}\xi^{\nu}\xi^{\lambda})$,
the action (\ref{2})
is invariant under (\ref{14}) and (\ref{15}) 
if one requires
\begin{equation}
\alpha(\tau_i)=\alpha(\tau_f)=0,\qquad
\beta(\tau_i)=\beta(\tau_f)=0.
\label{16}
\end{equation}
Formally ( $\beta\neq const$), 
the $\beta$-transformation is
the linear combination of the transformations
(\ref{12}) and (\ref{13}) with $m^{-1}\gamma=\omega$.
The $\beta$-transformation mixes the even coordinates $x_\mu$ with the
odd spin variables $\xi_\mu$; 
this transformation is
characterized
by an even parameter and, as we shall see in Section 3, it is generated
by an even (nilpotent) constraint.  As a consequence, one has
$\delta_{\beta_1}\delta_{\beta_2}-
\delta_{\beta_2}\delta_{\beta_1}=0$.

\subsection{Discrete symmetries}

The action is invariant under the discrete $P-$ and
$T-$tran\-sfor\-ma\-tions:
\begin{equation}
P:\, X_\mu\rightarrow
(X_0,-X_1,X_2),\qquad
T:\, X_\mu\rightarrow
(-X_0,X_1,X_2),
\label{17}
\end{equation}
where
$X_\mu=x_\mu,\xi_\mu,$ and
\begin{equation}
P:\, ,T:\, (e,v)\rightarrow
(e,-v),
\label{18}
\end{equation}
\begin{equation}
P:\, \theta_a\rightarrow
(\theta_1,-\theta_2).\qquad
T:\, \theta_a\rightarrow (-\theta_1,\theta_2).
\label{19}
\end{equation}
Since $P:\, \theta^\pm\rightarrow \theta^\mp$,
and $T:\, \theta^\pm\rightarrow -\theta^\mp$,
one has that $P:\,  ,T:\, {\cal N}\rightarrow -{\cal N}$.

Due to the global U(1) symmetry,
there is a freedom in the
choice of the form of the $P$ and $T$ transformations
in the sector of the variables $\theta_a$.

If one considers only the continuous and discrete symmetries 
described so far, the parameter $\nu$ can take any value in the 
classical theory.
As we shall see in Appendix A, there is a hidden classical
symmetry restricting the allowed values of $\nu$. This symmetry,
realized in the subspace of the variables $\theta^\pm$, is the
product of $P$ or $T$ (\ref{19}) with a special U(1)
transformation.

\section{Hamiltonian description}
\subsection{Equations and integrals of motion}

Let us turn to the canonical description of
the model and to the construction of the general solution to
the classical equations of motion. 

The nontrivial brackets of the system are the following:
\[
\{x_\mu,p_\nu\}=\eta_{\mu\nu},\quad
\{e,p_e\}=1,\quad
\{v,p_v\}=1,
\]
\begin{equation}
\{\xi_\mu,\xi_\nu\}=i\eta_{\mu\nu},\quad
\{\theta_a,\theta_b\}=-i\delta_{ab},
\label{20}
\end{equation}
and, hence,
\begin{equation}
\{\theta^+,\theta^-\}=-i.
\label{21}
\end{equation}
In correspondence with the two local symmetries
of the Lagrangian,
in the Hamiltonian formalism one has
two secondary first class constraints
\begin{equation}
\phi=\frac{1}{2}(p^2+m^2)\approx 0,
\label{22}
\end{equation}
\begin{equation}
\chi=i\epsilon_{\mu\nu\lambda}p^\mu\xi^\nu\xi^\lambda+2\nu m
{\cal N}\approx 0,
\label{23}
\end{equation}
which are the generators of the transformations (\ref{14}), (\ref{15})
and form the trivial algebra
\begin{equation}
\{\phi,\chi\}=0.
\label{24}
\end{equation}
In addition, there are two primary first class constraints
$p_e\approx 0$ and $p_v\approx 0$ 
merely stating the fact that $e$ and $v$ are
Lagrange multipliers. The Hamiltonian is a linear combination
of the constraints \cite{12}:
\begin{equation} H=e\phi+v\chi+u_1 p_e
+u_2 p_v,
\label{25}
\end{equation}
with
$u_{1,2}=u_{1,2}(\tau)$ arbitrary functions
of the evolution parameter $\tau$.

The equations of motion have the form:
\[
\dot{x}_\mu=ep_\mu+iv\epsilon_{\mu\nu\lambda}\xi^\nu
\xi^\lambda,\qquad
\dot{p}_\mu=0,
\]
\begin{equation}
\dot{\xi}_\mu=2v\epsilon_{\mu\nu\lambda}p^\nu\xi^\lambda,\qquad
\dot{\theta}{}^\pm=\pm 2i\nu mv\theta^\pm,
\label{26}
\end{equation}
and $\dot{e}=u_1$, $\dot{v}=u_2.$
From (\ref{26}), one identifies the set
of integrals of motion :
\begin{equation}
P_\mu=p_\mu,\qquad
J_\mu=-\epsilon_{\mu\nu\lambda}x^\nu p^\lambda-\frac{i}{2}
\epsilon_{\mu\nu\lambda}\xi^\nu\xi^\lambda,
\label{27}
\end{equation}
\begin{equation}
\Gamma=p\xi,\qquad
\Xi=i\epsilon_{\mu\nu\lambda}\xi^\mu\xi^\nu\xi^\lambda,\qquad
{\cal N}=\theta^+\theta^-,\qquad
\Delta=i\epsilon_{\mu\nu\lambda}p^\mu\xi^\nu\xi^\lambda.
\label{28}
\end{equation}
The integrals $P_\mu$ and $J_\mu$ are the generators of the
Poincar\'e transformations, whereas the integrals (\ref{28})
generate the global symmetries given by the transformations
(\ref{10})--(\ref{13}). The two last integrals of motion,
${\cal N}$ and $\Delta$, are related by
the constraint (\ref{23}) which is the generator of the
local $\beta-$transformations. This explains why 
the global symmetries (\ref{12}) and (\ref{13})
and the local symmetry (\ref{15}) are related.

Using the mass shell constraint, one may
introduce the complete oriented triad
$e^{(\alpha)}_{\mu}=e^{(\alpha)}_\mu(p)$, $\alpha=0,1,2$,
\begin{equation}
e^{(0)}_\mu=\frac{p_\mu}{\sqrt{-p^2}},\quad
e^{(\alpha)}_\mu\eta_{\alpha\beta}e^{(\beta)}_\nu=\eta_{\mu\nu},\quad
\epsilon_{\mu\nu\lambda}e^{(0)\mu}e^{(i)\nu}e^{(j)\lambda}=
\epsilon^{0ij}.
\label{30}
\end{equation}
Upon defining
$\xi^{(\alpha)}=\xi^{\mu}e^{(\alpha)}_\mu$
and
\begin{equation}
\xi^{(\pm)}=\frac{1}{\sqrt{2}}(\xi^{(1)}\pm i\xi^{(2)}),
\label{31}
\end{equation}
one has
\begin{equation}
\{\xi^{(+)},\xi^{(-)}\}=i,
\label{32}
\end{equation}
which differs in sign from the brackets
(\ref{21}) for the variables $\theta^\pm$.
This difference is relevant for the symmetries of the model.

The space-like components
of the triad $e^{(i)}_\mu$, $i=1,2,$ are not Lorentz vectors
(see, e.g. ref.  \cite{13}), and, therefore, the quantities
$\xi^{(i)}$ are not scalars. 

Taking
into account the mass shell constraint, one can present the
nilpotent constraint (\ref{23})
in the equivalent form:
\begin{equation}
{\cal X}= \xi^{(+)}\xi^{(-)}
-\nu\theta^+\theta^-\approx 0.
\label{33}
\end{equation}

With the help of the triad (\ref{30}) 
one finds yet another 
dependence among the integrals of motion
(\ref{27}), (\ref{28}).  Taking into account the
completeness of the triad, one gets the equality:
\begin{equation}
\Xi=3i(P^2)^{-1}\Gamma\Delta.
\label{34}
\end{equation}
As a result, one  has the
following combination of the integrals of motion:
\begin{equation}
\rho=\nu \cdot p\xi\theta^+\theta^--im\kappa\cdot
\epsilon_{\mu\nu\lambda}\xi^\mu\xi^\nu\xi^\lambda,
\label{35}
\end{equation}
which is an odd function and has the weakly vanishing bracket
\[
\{\rho,\rho\}=-\frac{3}{2}i\kappa\chi^2.
\]
Using (\ref{34}), on the surface of the
mass shell constraint, for $\kappa=1/6$, one
finds
\begin{equation}
\rho=-m\xi^{(0)}\cdot(\xi^{(+)}\xi^{(-)}-\nu\theta^+\theta^-).
\label{36}
\end{equation}
Thus, for $\kappa=1/6$, $\rho$
is proportional to $\chi$ and the additional constraint
\begin{equation}
\rho\approx 0
\label{37}
\end{equation}
can be imposed.

The classical odd
constraint
(\ref{37}) has the solution $\xi^{(0)}=0$ in addition
to the subspace singled out by the nilpotent constraint
(\ref{23}); consequently, its introduction modifies the physical content
of the original model.
At the quantum level, the analog of
$\xi^{(0)}$ is an invertible
operator; this implies that the quantum analog of the
constraint (\ref{36}) is equivalent to the quantum
analog of the constraint (\ref{23}).
In the quantum theory, the additional constraint (\ref{37}) 
does not change the physical content of the model. 
Of course, one may
exclude the point $\xi^{(0)}=0$ from the phase space of the 
classical system requiring that $\xi^{(0)}\neq 0$; in this case, 
the additional constraint 
$\rho\approx0$, for $\kappa=1/6$, does not change the physical content 
of the model even at the classical level.
From this one learns that
the physical content of the modified classical model
--- i.e. the model supplied with the constraint (\ref{35}), (\ref{37})
--- is governed by the $c-$number parameter $\kappa$ 
and that, for  the particular value $\kappa=1/6$, the additional
odd constraint differs from the even nilpotent constraint
(\ref{23}) only in the factor $\xi^{(0)}$. Hence, one may change
Grassmann parity of the integrals of motion multiplying them by
the odd integral $\xi^{(0)}$.  These observations are helpful
for unveiling the hidden continuous symmetries of the model.

\subsection{Solution to the equations and additional integrals
of motion}

Let us go back to the discussion of our original model
defined by the
constraints  (\ref{22}), (\ref{23}) and by the Hamiltonian
(\ref{25}).

Since $\xi^{(0)}=const$, the
equations of motion for the odd variables
\begin{equation}
\xi_\mu=
-\xi^{(0)}e^{(0)}_\mu+\xi^{(i)}e^{(i)}_\mu
\label{38}
\end{equation}
have the same form
(up to the constant $\nu$) as the
equations for the variables $\theta^{\pm}$. Namely,
\begin{equation}
\dot{\xi}{}^{(\pm)}=\pm 2imv\xi^{(\pm)}.
\label{39}
\end{equation}
{}From (\ref{39}) and (\ref{26}) 
one concludes that, if 
\begin{equation}
\nu=1,
\label{40}
\end{equation}
the odd variables $\theta^\pm$ and $\xi^{(\pm)}$ have
a harmonic-like evolution law with equal
(in general time-dependent) angular velocities.
As a result, for $\nu=1$
one has two additional
mutually conjugate integrals of motion:
\begin{equation}
{\cal V}_+=i\xi^{(+)}\theta^-,\qquad
{\cal V}_-={\cal V}_+^*=i\xi^{(-)}\theta^+,
\label{41}
\end{equation}
which generate the global symmetries of the
model.

Except when explicitly stated otherwise, in the following 
we shall fix $\nu=1$.
For this value of the parameter the model is maximally
symmetric, i.e. it has the maximal number of integrals of motion.

The  general solution to the equations of motion is given
by 
\begin{equation}
\xi^{(\pm)}(\tau)=\xi^{(\pm)}(\tau_i)e^{\pm
i\Omega(\tau;\tau_i)},\qquad
\theta^\pm(\tau)=\theta^\pm(\tau_i)e^{\pm i\Omega(\tau;\tau_i)},
\label{42}
\end{equation}
\begin{equation}
x_\mu(\tau)=p_\mu\cdot\int_{\tau_i}^{\tau}e(\tau')d\tau'
+e^{(i)}_\mu \left(x^{(i)}_Z(\tau) -
x^{(i)}_{Z}(\tau_i)\right)
+x_\mu(\tau_i),
\label{43}
\end{equation}
\begin{equation}
x^{(i)}_Z(\tau)=-im^{-1}\xi^{(0)}\xi^{(i)}(\tau).
\label{44}
\end{equation}
$\Omega(\tau;\tau_i)$ is defined by the relation
\begin{equation}
\Omega(\tau;\tau_i)=2m\int_{\tau_i}^{\tau}v(\tau')d\tau'.
\label{45}
\end{equation}
We assume that the constraints (\ref{22}), (\ref{23}) are
satisfied by $p_\mu$ and by the initial values of the variables
$\theta^\pm$ and $\xi^{(\pm)}$, i.e. by $\theta^{\pm}(\tau_i)$
and $\xi^{(\pm)}(\tau_i)$ \cite{12}.
In writing the solution to the equations of
motion for the `oscillating' odd variables $\theta^\pm$ and
$\xi^{(\pm)}$, the pertinent boundary
conditions (\ref{7}) have been taken into account.  
This point is discussed in detail in
the Appendix A.

Eq.  (\ref{44}) describes the classical
analog of the quantum Zitterbewegung \cite{13}.
From eqs. (\ref{43}) and
(\ref{44}) one finds that
\[
Z^i=x^{(i)}+im^{-1}\xi^{(0)}\xi^{(i)}
\]
are integrals of motion.

The evolution of the odd vector variable $\xi_\mu$ is
written in covariant form as
$
\xi_\mu(\tau)=G_{\mu\nu}(\tau,\tau_i)\xi^\nu(\tau_i),
$
with $G_{\mu\nu}(\tau,\tau_i)$ given by
\[
G_{\mu,\nu}(\tau,\tau_i)=-e^{(0)}_\mu e^{(0)}_\nu
+\left(\eta_{\mu\nu}+e^{(0)}_\mu e^{(0)}_\nu\right)
\cos\Omega(\tau;\tau_i)
+\epsilon_{\mu\nu\lambda}e^{(0)\lambda}\sin\Omega(\tau;\tau_i).
\]

When $\nu\neq 1$
one may construct the integrals of motion:
\[
{\cal V}_{\nu+}\equiv i\xi^{(+)}\theta^- \exp (i(\nu-1)\Omega(\tau;\tau_i)),
\quad
{\cal V}_{\nu-}={\cal V}_{\nu+}^*,
\]
which are {\it nonlocal
(in $\tau$) functions} due to the presence of the
factor $\exp(i(\nu-1)\Omega(\tau;\tau_i))$.
They become 
local integrals
(\ref{41}) when $\nu=1$.

\subsection{Global U(1,1)  symmetry and $N=3$ supersymmetry}

Let us consider the following linear combinations of the
nilpotent even integrals  of motion ${\cal V}_+$, ${\cal V}_-$,
$\Delta$ and ${\cal N}$:
\[
R_{0}=-\frac{1}{2}(\xi^{(+)}\xi^{(-)}+\theta^+\theta^-),
\]
\begin{equation}
R_1=\frac{i}{2}(\xi^{(+)}\theta^-+\xi^{(-)}\theta^+),\quad
R_2=\frac{1}{2}(\xi^{(+)}\theta^--\xi^{(-)}\theta^+).
\label{46}
\end{equation}
They are real quantities,
$R^*_\alpha=R_\alpha$, and
satisfy the $su(1,1)$ algebra
\begin{equation}
\{R_\alpha,R_\beta\}=-\epsilon_{\alpha\beta\gamma}R^\gamma.
\label{47}
\end{equation}
Moreover, they are in involution with
the nilpotent constraint
\begin{equation}
{\cal X}=\xi^{(+)}\xi^{(-)}-\theta^+\theta^-\approx 0,
\label{48}
\end{equation}
since $\{R_\alpha,{\cal X}\}=0$. Therefore,
the set of integrals of motion $R_\alpha$ and ${\cal X}$
form a $u(1,1)=su(1,1)\times u(1)$ algebra.
The  algebra of the pseudounitary group
U(1,1) appears since the pairs
of variables $\theta^\pm$ and $\xi^{(\pm)}$ have 
brackets differing in sign.
Due to eq. (\ref{26}),
${\cal X}$ plays the role of the Hamiltonian 
for the odd variables $\xi_\mu$ and $\theta^\pm$.

The Casimir operator of the group SU(1,1),
$C=R_\alpha R^\alpha$,
\begin{equation}
\{C,R_\alpha\}=0,
\label{49}
\end{equation}
takes the value
\[
C=-\frac{3}{2}\xi^{(+)}\xi^{(-)}\theta^+\theta^-\approx 0,
\]
and it is related to $R_\alpha$ through 
the classical equation:
\begin{equation}
R_\alpha R_\beta=\eta_{\alpha\beta}\cdot\frac{1}{3}C.
\label{50}
\end{equation}

The integral $R_0$
can be written in the manifestly covariant form
\[
R_0= \frac{1}{2}\left(\frac{i}{2\sqrt{-p^2}}
\epsilon_{\nu\rho\lambda}p^\nu\xi^\rho\xi^\lambda-
\theta^+\theta^-\right),
\]
and the integrals $R_{1,2}$ can be obtained from 
the Lorentz vector
\begin{equation}
R^{\perp}_{\mu}=
\frac{i}{2}\left(\xi_\mu-p_\mu
\frac{p\xi}{p^2}\right)\theta_1
-\frac{i}{2\sqrt{-p^2}}\epsilon_{\mu\nu\lambda}p^\nu\xi^\lambda
\theta_2,
\label{51}
\end{equation}
which satisfies the relations $R_\mu^\perp p^\mu=0$ and
$R_{\mu}^\perp e^{(i)\mu}=R_i$.
One can now construct the vector
\begin{equation}
{\cal R}_\mu=R_{\mu}^\perp + e^{(0)}_\mu R_0,
\label{52}
\end{equation}
providing the covariant set of SU(1,1) generators, which,
due to ${\cal R}^\mu e^{(\alpha)}_\mu=R^\alpha$,
is equivalent to the noncovariant set (\ref{46}).

Let us consider the following odd integrals of motion,
which are the composition of the integrals (\ref{52}) and
$\xi^{(0)}$:
\begin{equation}
\Theta_\mu=\xi^{(0)}{\cal R}_\mu.
\label{54}
\end{equation}
Due to (\ref{50}), the odd integrals (\ref{54})
together with the even
integral $C$ satisfy the 
covariant algebra of an $N=3$ supersymmetry:
\begin{equation}
\{\Theta_\mu,\Theta_\nu\}=-\eta_{\mu\nu}\cdot\frac{i}{3}C,\qquad
\{\Theta_\mu,C\}=0.
\label{55}
\end{equation}
The relation between the even
integrals, ${\cal R}_\mu$, and the odd ones, $\Theta_\mu$, is
similar to the one existing between the constraint  $\chi$ and
the additional constraint $\rho$. 

In this section we
have unveiled the global U(1,1) 
symmetry and the $N=3$ supersymmetry 
of the classical model.
The even generator of the supersymmetry, $C$, differs from the
constraint ${\cal X}$ which is, in fact, a Hamiltonian for all
the Grassmann variables of the system.  This non-standard
supersymmetry is analogous to that discovered in ref. \cite{1}.

\subsection{Covariant form of the U(1,1) symmetry}

With the help of the standard prescription \cite{12}
one finds the form of the 
global transformations generated by the even,
${\cal R}_\mu$, and odd, $\Theta_\mu$, integrals of motion:
\begin{equation}
\delta_{\omega_\mu}X=\omega_\mu\{X,{\cal R}^\mu\},
\quad
\delta_{\epsilon_\mu}X=\epsilon_\mu\{X,\Theta^\mu\},\quad
X=x_\mu,\xi_\mu,\theta_a.
\label{53}
\end{equation}
In (\ref{53}) $\omega_\mu$ and $\epsilon_\mu$
are even and odd constant infinitesimal transformation
parameters, respectively.
Both transformations (\ref{53})
mix the even coordinates $x_\mu$ with the `internal'
translation-invariant odd variables of the model.
Since the explicit form of the transformation properties of
the coordinates $x_\mu$  is not needed,
we shall not write it here.

A more detailed analysis
of the transformations (\ref{53})
for the odd variables gives the possibility to `covariantize'
the U(1,1) symmetry of the theory.
For this purpose, one introduces the following linear combinations
$\Sigma^{i}_a$, $a=1,2$, $i=1,2$,
of the odd variables
$\xi^{(i)}$ and $\theta_a$,
\[
\Sigma^1_1=\xi^{(1)}+\theta_1,\quad
\Sigma^1_2=-\xi^{(2)}+\theta_2,
\]
\begin{equation}
\Sigma^2_1=\xi^{(2)}+\theta_2,\quad
\Sigma^2_2=\xi^{(1)}-\theta_1.
\label{spinor}
\end{equation}
These variables are spinors (in the index $a$)
with respect to the action of the 
SU(1,1) generators $R_\alpha$ and
form SO(2) vectors (in the index $i$) 
with respect to the action of the generator ${\cal X}$.This 
gives the possibility 
to `covariantize' the SU(1,1) generators $R_\alpha$ as well
as the SO(2) (or U(1)) generator ${\cal X}$.
In fact, let us introduce the
SO(2) tensor $\epsilon^{ij}=-\epsilon^{ji}$ and the SU(1,1)
spinor tensor
$\tilde{\epsilon}{}^{ab}=-\tilde{\epsilon}{}^{ba}$, as well as
its inverse, $\tilde{\epsilon}_{ab}=-\tilde{\epsilon}_{ba}$,
normalized as $\epsilon^{12}=\tilde{\epsilon}^{12}=
\tilde{\epsilon}_{12}=1$, and 
let us define $\Sigma^{ia}=\tilde{\epsilon}{}^{ab}\Sigma^{i}_{b}$.
Lowering and rasing the `spinor' indices
is understood only for the variables $\Sigma^{i}_a$,
but not for $\xi^{(1,2)}$ and $\theta_{1,2}$.
The new variables have the following brackets:
\begin{equation}
\{\Sigma^{ia},\Sigma^{j}_b\}=-2i\epsilon^{ij}\delta^a_b.
\label{ssd}
\end{equation} 
In terms of $\Sigma^i_a$ 
the U(1,1) generators are written in the 
`covariant' form,
\begin{equation}
R_\alpha=-\frac{1}{8}\epsilon^{ij}\Sigma^{ia}(\tilde{\gamma}_\alpha)_a{}^b
\Sigma^{j}_b,
\label{rss*}
\end{equation}
\begin{equation}
{\cal X}=-\frac{i}{4}\Sigma^{ia}\Sigma^{i}_a.
\label{css*}
\end{equation}
In eqs. (\ref{rss*})
we have introduced the $\gamma-$matrices in the Majorana
representation,
$(\tilde{\gamma}{}^0)_a{}^b=-(\sigma^2)_a{}^b$,
$(\tilde{\gamma}{}^1)_a{}^b=i(\sigma^3)_a{}^b$,
$(\tilde{\gamma}{}^2)_a{}^b=-i(\sigma^1)_a{}^b$,
$\tilde{\gamma}_\alpha\tilde{\gamma}_\beta=
-\eta_{\alpha\beta}+i\epsilon_{\alpha\beta\gamma}
\tilde{\gamma}{}^{\gamma}$,
and we used the tilde to stress that they are constant 
in the tangent space and differ from the matrices 
$\gamma^{(\alpha)}=e^{(\alpha)}_\mu\gamma^\mu$.
The variables $\Sigma^i_a$ are indeed
SU(1,1) spinors and SO(2) vectors:
\begin{equation}
\{\Sigma^i_b,R_\alpha\}=\frac{i}{2}
(\tilde{\gamma}_\alpha)_a{}^b\Sigma^i_b,
\label{rss}
\end{equation}
\begin{equation}
\{\Sigma^i_a,{\cal X}\}=\epsilon^{ij}\Sigma^j_a.
\label{css}
\end{equation} 
Defining the even variables
\begin{equation}
\Xi^i_a=i\xi^{(0)}\Sigma^i_a,
\label{xis}
\end{equation}
one readily finds that they satisfy to
\begin{equation}
\{\Xi^i_b,R_\alpha\}=\frac{i}{2}
(\tilde{\gamma}_\alpha)_a{}^b\Xi^i_b,
\label{rxx}
\end{equation}
\begin{equation}
\{\Xi^i_a,{\cal X}\}=\epsilon^{ij}\Xi^j_a.
\label{cxx}
\end{equation} 
Taking into account that
\[
i\{\Xi^i_a,\Xi^j_b\}=\Sigma^i_a\Sigma^j_b,
\]
one gets that (\ref{rss*}) and (\ref{css*})
can be written as:
\begin{equation}
R_\alpha=-\frac{i}{8}\epsilon^{ij}(\tilde{\gamma}_\alpha)_a{}^b
\{\Xi^{ia},\Xi^j_b\},
\label{xixir}
\end{equation}
\begin{equation}
{\cal X}=\frac{1}{4}\{\Xi^{ia},\Xi^i_a\}.
\label{xixich}
\end{equation}
Eqs. (\ref{rxx})--(\ref{xixich}) 
express covariantly 
the U(1,1) symmetry of the system.

\section{Gauging the U(1) symmetry}

The action is invariant under the global U(1)
transformations (\ref{12}) generated by the integral of motion
${\cal N}$:
\begin{equation}
\theta^\pm\rightarrow \theta^\pm{}'=e^{\pm i\gamma}\theta^\pm.
\label{56}
\end{equation}
Let us assume that $\gamma$ is not a constant, i.e.
$\dot{\gamma}\neq 0$.
Since under (\ref{56})   one has 
$\delta_\gamma L_\theta=d\gamma/d\tau\cdot {\cal N}$,
the lagrangian of the system is not invariant.
In addition, if one assumes that
$\gamma=\gamma(x_\mu(\tau))$,
one has
$\delta_\gamma L=\partial_\mu\gamma
\cdot\dot{x}{}^\mu {\cal N}$.
Upon introducing a `compensating'
term 
\begin{equation}
L^1_{int}={\cal Q}{\cal N}\dot{x}{}^\mu
{\cal A}_\mu
\label{57}
\end{equation}
in the Lagrangian, with 
${\cal Q}$ 
a coupling constant
and 
${\cal A}_\mu={\cal A}_\mu(x)$ 
a gauge field 
transforming as
\begin{equation}
\delta_\gamma {\cal A}_\mu=-{\cal Q}^{-1}\partial_\mu\gamma(x),
\label{58}
\end{equation}
one easily verifies that
$\delta_\gamma (L+L^1_{int})=0$. The term
$L^1_{int}$ does not change the reparametrization invariance, but 
it violates the $\beta-$invariance of the
action. To restore the latter invariance, one introduces the
additional term
\begin{equation}
L^2_{int}=\frac{e}{2}{\cal Q}{\cal N}i\xi_\mu\xi_\nu {\cal
F}^{\mu\nu},\qquad
{\cal F}_{\mu\nu}=\partial_\mu {\cal
A}_\nu-\partial_\nu {\cal A}_\mu,
\label{59}
\end{equation}
which is 
invariant under the U(1) gauge
transformations (\ref{56}), (\ref{58}).  Since the 
Lagrangian
\begin{equation}
L_{int}=L^1_{int}+L^2_{int}={\cal Q}{\cal
N}\left(\dot{x}_\mu {\cal A}^\mu +\frac{e}{2}i\xi_\mu\xi_\nu
{\cal F}^{\mu\nu}\right)
\label{li}
\end{equation}
satisfies the condition that $\delta_\beta
L_{int}={\cal Q}\frac{d}{d\tau} ({\cal N}\delta_\beta x_\mu
{\cal A}^\mu)$,
$L+L_{int}$ is invariant under the local U(1) and $\beta-$ symmetries.

The boundary term $B_\theta$ is invariant
under the local U(1) transformations if the
function $\gamma(x)$ satisfies the boundary condition
\begin{equation}
\Delta\gamma=\gamma(x(\tau_f))-\gamma(x(\tau_i))
=2\pi n,\quad n\in Z.
\label{60}
\end{equation}

The U(1) gauge
interaction preserves the invariance of the action under the 
discrete $P$ and $T$ inversions if one
supplements the transformation laws (\ref{17})--(\ref{19})
with 
\begin{equation}
P:\, {\cal A}_\mu\rightarrow (-{\cal A}_0,{\cal
A}_1,-{\cal A}_2),\quad
T:\, {\cal A}_\mu\rightarrow ({\cal A}_0,-{\cal
A}_1,-{\cal A}_2).
\label{61}
\end{equation}

In addition to the 
U(1) gauge symmetry described above,
one may introduce the usual electromagnetic interaction
\begin{equation}
L_{int}^{em}=q\left(\dot{x}_\mu A^\mu+\frac{1}{2}
ei\xi_\mu\xi_\nu F^{\mu\nu}\right)
\label{liem}
\end{equation}
with $F_{\mu\nu}=\partial_\mu A_\nu-\partial_\nu A_\mu$.
Under $P$ and $T$ inversions the
electromagnetic potential transforms as
\[
P:\, A_{\mu}\rightarrow
(A_{0},-A_{1},A_{2}),\quad T:\, A_{\mu}\rightarrow
(-A_{0},A_{1}, A_{2}).
\]
It is the difference in the
transformation law of the gauge potentials ${\cal A}_\mu(x)$
and $A_\mu(x)$ under the
discrete $P$ and $T$ inversions what distinguishes the
two types of U(1) gauge interactions.

At the Hamiltonian level the first class constraints (\ref{22}),
(\ref{23}) are now modified as
\begin{equation}
\tilde{\phi}=\frac{1}{2}\left({\cal
P}^2+m^2-i\xi_\mu\xi_\nu
({\cal QN}{\cal F}^{\mu\nu}+
qF^{\mu\nu})\right)\approx0,
\label{62}
\end{equation}
\begin{equation}
\tilde{\chi}=i\epsilon_{\mu\nu\lambda}{\cal
P}^{\mu}\xi^\nu\xi^\lambda-2m{\cal N}\approx0,
\label{63}
\end{equation}
where
\[
{\cal P}_\mu=p_\mu-{\cal Q}{\cal N}{\cal A}_\mu-qA_{\mu}.
\]
These constraints satisfy the same algebra  of the first class
constraints of the free model, i.e.
$\{\tilde{\phi},\tilde{\chi}\}=0$.

To conclude the analysis of the classical theory, let us
comment on a property 
of the U(1) gauge symmetry
(\ref{56}), (\ref{58}), (\ref{60}).
As it is shown in Appendix~A, 
there is a hidden symmetry in the free theory
(${\cal A}_\mu=A_\mu=0$),
which is realized only
in the sector of the variables $\theta_a$.
This symmetry leads
to a `quantization' condition on the parameter $\nu$
(here the maximal symmetry
of the classical system is not required, thus $\nu$ can take any value).
Namely, 
$\nu=\nu_{lk}$,
\begin{equation}
\nu_{lk}=\frac{2l+1}{2k+1},\quad
k,l\in Z.
\label{64}
\end{equation}
This value of the parameter defines the ratio
of the phase changes of the harmonic-like variables
$\theta^\pm$ and $\xi^{(\pm)}$ for the complete time interval
$\Delta\tau=\tau_f-\tau_i$ (see eq. (\ref{87})).
As we have seen in section 3, the
requirement of the maximal symmetry in the system
selects from the infinite set (\ref{64})
only the value
$\nu=1$. As a result, according to eq. (\ref{26}),
the phases of
the harmonic-like variables $\xi^{(\pm)}$ and $\theta^\pm$ turn
out to be locked (see eq. (\ref{42}));
in fact, the ratio of the phase changes
for the complete time interval should be equal to 1.
After switching on the interaction
with the U(1) gauge field ${\cal A}_\mu$,
the effect of the finite gauge
transformation of the variables $\theta^\pm$,
given by eq. (\ref{56}) with the
gauge function $\gamma(x)$  subject
to the condition (\ref{60}), 
only restores 
the initial freedom of the ratio of the phase changes of the
harmonic-like variables.
In fact,
according to eqs.  (\ref{85}), (\ref{87}),
(\ref{56}) and (\ref{60}), for
$\nu=1$ one has
\begin{equation}
\theta^{\pm}{}'(\tau_f)=e^{\pm
i(\Omega(\tau_f)+2\pi n)}\theta^{\pm}{}'(\tau_i).
\label{65}
\end{equation}
As a result, the ratio of the phase changes of
the transformed variables $\theta^\pm{}'$ and of the
variables $\xi^{(\pm)}$, given by (\ref{65}),
(\ref{85}) and (\ref{87}),
is described again by (\ref{64}) with
$l=k+n$.

\section{Quantization of the model}
\subsection{Free theory}
Due to (\ref{20}),
the quantum operators associated with the odd variables must satisfy
the following anticommutation relations:
\[
[\hat{\xi}_\mu,\hat{\xi}_\nu]_{{}_+}=-\eta_{\mu\nu},\qquad
[\hat{\theta}_a,\hat{\theta}_b]_{{}_+}=\delta_{ab},\quad
[\hat{\xi}_\mu,\hat{\theta}_a]_{{}_+}=0.
\]
Here and below we put $\hbar=1$.
The operators $\hat{\xi}{}^{\mu}$ and
$\hat{\theta}_a$ are realized as
\[
\hat{\xi}{}^\mu=\frac{1}{\sqrt{2}}\gamma^\mu\otimes
\sigma_3,\qquad
\hat{\theta}_a=\frac{1}{\sqrt{2}}1\otimes\sigma_a,\qquad
a=1,2.
\]
It is convenient to take the $\gamma$-matrices, 
satisfying the relation
\[
\gamma_\mu\gamma_\nu=-\eta_{\mu\nu}+i\epsilon_{\mu\nu\lambda}
\gamma^\lambda,
\]
in the form
$\gamma^0=\sigma_3$, $\gamma^i=i\sigma^i$, $i=1,2$.
The classical first class constraints (\ref{22}), (\ref{23})
turn into  
quantum equations singling out the physical subspace of the
system:
\begin{equation}
\hat{\phi}\Psi=0,
\label{66}
\end{equation}
\begin{equation}
\hat{\chi}\Psi=0.
\label{67}
\end{equation}
In the construction of the quantum operator corresponding to the nilpotent
constraint 
one has an ordering problem. 
For arbitrary $\nu$, $\hat{\chi}$ is written as
\begin{equation}
\hat{\chi}=p\gamma+2\nu m\hat{{\cal N}}.
\label{68}
\end{equation}
If one starts with
the classical expression
${\cal N}=\theta^+\theta^-$, 
one has \cite{9}:
\[
\hat{{\cal N}}=\alpha\hat{\theta}{}^+\hat{\theta}{}^-+(\alpha-1)
\hat{\theta}{}^-\hat{\theta}{}^+,
\]
with $\alpha$ an arbitrary real parameter.
For the sake of simplicity, in eq. (\ref{68})
and below the operator of the energy-momentum vector
of the system has been denoted in the same way as its classical counterpart.
Upon multiplying the second quantum equation
(\ref{67}) with the operator $p\gamma-2\nu m\hat{{\cal N}}$,
one gets
\[ 
(p^2+4\nu^2 m^2
\hat{{\cal N}}{}^2)\Psi=0,
\]
which reduces to the mass shell
condition (\ref{66}) (Klein-Gordon equation) iff
\begin{equation}
4\nu^2\hat{{\cal N}}{}^2=1.
\label{69}
\end{equation}
Eq. (\ref{69}) is satisfied only when
\[
\alpha=\frac{1}{2},\qquad \nu^2=1.
\]
One immediately concludes
that the choice (\ref{40}) is a special one also in 
the quantum theory. The value
$\alpha=1/2$ leads to the operator
\begin{equation}
\hat{{\cal N}}=\frac{1}{2}[\hat{\theta}{}^+,\hat{\theta}{}^-]=-i
\hat{\theta}_1\hat{\theta}_2=\frac{1}{2}\cdot 1\otimes\sigma_3.
\label{70}
\end{equation}
Eq. (\ref{67}) takes the form
\begin{equation}
(p\gamma\otimes 1+m\cdot 1\otimes\sigma_3)\Psi=0.
\label{71}
\end{equation}
Therefore, the components $\psi_u$ and $\psi_d$ of the function
$\Psi$, $\Psi^t=(\psi_u^t,\psi_d^t)$, satisfy a pair 
of (2+1)-dimensional Dirac equations:
\begin{equation}
(p\gamma+m)\psi_u=0,\qquad
(p\gamma-m)\psi_d=0.
\label{72}
\end{equation}
In correspondence with the definition of the
(2+1)-dimensional spin operator,
\[
\hat{S}=\frac{p\hat{J}}{\sqrt{-p^2}},
\]
where $\hat{J}_\mu$ is the quantum operator corresponding to the total angular
momentum of the system defined in eq. (\ref{27}),
one has
\begin{equation}
\hat{S}=-\frac{1}{2}\cdot \gamma^{(0)}\otimes 1.
\label{73}
\end{equation}
Due to eqs. (\ref{72}), the physical
states $\psi_u$ and $\psi_d$ carry spin $+1/2$ and $-1/2$,
respectively.

A  $P-$ and $T-$invariant
quantum system emerges from the classical picture.
In fact, 
starting from the classical relations (\ref{17})--(\ref{19}),
one finds the corresponding quantum transformation laws:
\begin{equation}
P:\, \Psi(x)\rightarrow
\Psi'(x')=(\gamma^1\otimes\sigma_1)\cdot\Psi(x),\quad
x'_\mu=(x_0,-x_1,x_2),
\label{74}
\end{equation}
\begin{equation}
T:\, \Psi(x)\rightarrow
\Psi'(x')=(\gamma^0\otimes\sigma_2)\cdot\Psi(x),
\quad
x'_\mu=(-x_0,x_1,x_2),
\label{75}
\end{equation}
By direct inspection one easily verifies that
eq. (\ref{71}) is invariant with respect to these transformations.

Of course, the quantum system has also the charge conjugation symmetry
\begin{equation}
C:\, \Psi(x)\rightarrow
\Psi_c(x)=(\sigma^1\otimes 1)\cdot\Psi^*(x)\ .
\label{75c}
\end{equation}
When the quantum system is interacting with
the Abelian gauge fields $A_\mu(x)$ and ${\cal A}_\mu(x)$, under the 
charge conjugation symmetry one has to
require that both such fields change sign.

Note that the system of equations (\ref{67}) has
also nontrivial solutions when $\nu\neq 1$ and $\alpha$ satisfy
either the condition $(2\nu\alpha)^2=1$ or the condition
$(2\nu(\alpha-1))^2$.
In these two cases one has only one nontrivial state given by
the function $\psi_u$ or $\psi_d$, respectively. 
The $P-$ and $T-$ invariance of
the classical action is lost.\footnote{
The discrete $P-$ and $T-$symmetries
were not discussed at the classical level in ref. \cite{9}
since it was assumed that the variables 
$\theta_a$ and $v$ were scalars.}.
For $\nu\neq 1$ one has an anomalous quantum scheme.
For $\nu=1$,
one gets
\begin{equation}
\hat{\chi}^2=-2\hat{\phi}+4m \hat{{\cal N}}\hat{\chi},
\label{chif}
\end{equation}
i.e. at the quantum level, unlike in the classical theory,
the mass shell operator $\hat{\phi}$ and the operator $\hat{\chi}$
are dependent operators.

\subsection{Super-extension of the Poincar\'e group}

The quantum counterparts of the integrals (\ref{46}) are:
\begin{eqnarray}
&\hat{R}_0=\frac{1}{4}
(\gamma^{(0)}\otimes
1-1\otimes\sigma_3),&\nonumber\\
&\hat{R}_1=\frac{1}{4}(\gamma^{(2)}\otimes\sigma_1-
\gamma^{(1)}\otimes\sigma_2),\quad
\hat{R}_2=-\frac{1}{4}(\gamma^{(1)}\otimes\sigma_1
+\gamma^{(2)}\otimes\sigma_2).&
\label{76}
\end{eqnarray}
They satisfy the commutation relations of the $su(1,1)$ algebra
\begin{equation}
[\hat{R}_\alpha,\hat{R}_\beta]=-i
\epsilon_{\alpha\beta\gamma}
\hat{R}{}^{\gamma}.
\label{77}
\end{equation}
In the quantum theory
\[
\hat{C}=\hat{R}_\alpha\hat{R}{}^\alpha=\frac{3}{8}\left(\gamma^{(0)}
\otimes\sigma_3 -1\right)
\]
is the Casimir operator, which
can be written as
\begin{equation}
\hat{C}=\frac{3}{4}
\left((\hat{\cal N}-\hat{S})^2-1\right).
\label{cns}
\end{equation}
On the physical subspace it is reduced to
$\hat{C}=-3/4$
(or $C=-\hbar^2\cdot 3/4$ in units of $\hbar$).
Hence, the operators
$\hat{R}_\alpha$, forming the $su(1,1)$ algebra (\ref{77}), act 
irreducibly on the physical subspace.  The physical
states $\psi_u$ and $\psi_d$ are the eigenstates of the
operator
\[
\hat{R}_0=-\frac{1}{2}(\hat{{\cal N}}+\hat{S})
\]
with the eigenvalues $-1/2$ and $+1/2$, respectively.

Since at the
quantum level one has that:
\begin{equation}
\hat{\xi}{}^{(0)}\hat{R}_\alpha=-\frac{1}{\sqrt{2}}\hat{R}_\alpha,
\label{78}
\end{equation}
the generators of the SU(1,1)
symmetry, $\hat{R}_\alpha$, coincide up to a $c-$number factor 
with the quantum operators corresponding to the $N=3$ SUSY generators 
(\ref{54}).
Therefore, the operators $\hat{R}_\alpha$ satisfy 
not only the algebra (\ref{77}),  
but also the $s(3)$ superalgebra \cite{s3}
\begin{equation} 
[\hat{R}_\alpha,\hat{R}_\beta]_{{}_+}=\eta_{\alpha\beta}
\cdot\frac{2}{3}\hat{C},\qquad
[\hat{R}_\alpha,\hat{C}]=0.
\label{79}
\end{equation}
This corresponds
to the classical algebra (\ref{61})
of the SUSY generators (\ref{59}), (\ref{60}).
Due to relation similar in form to (\ref{78}), the
quantum operator corresponding to the 
odd function $\rho$ given by eq. (\ref{36})
coincides (up to a $c-$number factor) with the quantum constraint
$\hat{\chi}$.

Before going over to the analysis of the interacting
quantum system,
let us comment on the
hidden global (super)symmetry.
Upon writing down the 
quantum counterpart of the classical vector
(\ref{52}) as
$\hat{\cal R}_\mu=e^{(0)}_\mu\hat{R}_0+
\hat{R}_\mu^\perp$,
one sees that this operator satisfies the (anti)commutation 
relations
\begin{equation}
\hat{\cal R}_\mu\hat{\cal R}_\nu=
\frac{1}{3}\eta_{\mu\nu}\hat{C}
-\frac{i}{2}\epsilon_{\mu\nu\lambda}\hat{\cal R}{}^\lambda,
\quad
[\hat{\cal R}_\mu,\hat{C}]=0.
\label{80}
\end{equation}
Since $\hat{\cal R}_\mu$ is a Lorentz vector
operator,
\[
[\hat{J}_\mu,\hat{\cal R}_\nu]=-i\epsilon_{\mu\nu\lambda}
\hat{\cal R}{}^{\lambda},
\]
and it commutes with the
operator $p_\mu$,
\[
[\hat{\cal R}_\mu,p_\nu]=0,
\]
the set of vector operators $\hat{\cal R}_\mu$ and $p_\mu$
supplied with the total angular momentum operator $\hat{J}_\mu$
form a super-extension of the Poincar\'e 
group, denoted as ISO(2,1$\vert$3).
The
vector operator
\[
\hat{\cal J}_\mu=\hat{J}_\mu-\hat{\cal R}_\mu
\]
satisfies the same $su(1,1)$ algebra of the operators
$\hat{J}_\mu$ and $\hat{\cal R}_\mu$; namely,
\[
[\hat{\cal J}_\mu,\hat{\cal J}_\nu]=-i
\epsilon_{\mu\nu\lambda}\hat{\cal J}{}^{\lambda}.
\]
The Casimir operators of the supergroup
ISO(2,1$\vert$3) are the operators
$p^2$
and $\hat{\cal S}=e^{(0)}_\mu\hat{\cal J}{}^\mu$.
It is natural to name the operator
\begin{equation}
\hat{\cal S}=\frac{1}{2}(\hat{S}-\hat{{\cal N}})
\label{*s}
\end{equation}
the {\it superspin}.
The operator
$\hat{C}=\hat{\cal R}_\mu\hat{\cal R}{}^\mu$
can be written as a quadratic function of the superspin,
\[
\hat{C}=3\hat{\cal S}^2-\frac{3}{4}.
\]
Taking into account the explicit form of the operators
$\hat{\cal N}$ and $\hat{S}$ given by eqs. (\ref{70})
and (\ref{73}), one finds that the complete set of the eigenvalues of
the superspin is given by the numbers $(-1/2,0,0,+1/2)$.
The physical states $\psi_u$ and $\psi_d$,
carrying spin $+1/2$ and $-1/2$,
are simultaneously  the eigenstates of the superspin operator
with zero eigenvalue, 
i.e. the one particle states of the quantum 
$P$, $T$---invariant theory
realize an irreducible representation of the supergroup
ISO(2,1$\vert$3) labelled by the zero eigenvalue of 
the superspin.

The nonstandard character of the super-extension
of the Poincar\'e group is related to the vector nature of
the supercharge operators $\hat{\cal R}_\mu$,
satisfying also the commutation relations
of the SU(1,1) generators.

In conclusion, we explicitly constructed 
the nontrivial `superposition' of the
discrete ($P$ and $T$) and continuous ($U(1,1)$ and $S(3)$)
(super)symmetries characteristic of the $P,T-$ invariant planar fermion model.
The generators of the continuous (super)symmetries are 
combined with the Poincar\'e generators resulting in the 
non-standard superextension of the Poincar\'e group.

\subsection{Interacting quantum theory}

In constructing the quantum operators corresponding to the first
class constraints of the classical theory
interacting with the U(1) gauge field ${\cal A}_\mu$
one should recall the form (\ref{70}) of  
the operator $\hat{{\cal N}}$ since
the physical states $\psi_u$ and
$\psi_d$, carrying opposite spins $+1/2$ and $-1/2$, are
distinguished by this operator; in fact,
these states carry opposite U(1) charges $+{\cal Q}/2$ and $-{\cal Q}/2$.
Since,
under $P$ and $T$ inversions, one has $\hat{\cal N}\rightarrow
-\hat{\cal N}$, the physical states
$\psi_u$ and $\psi_d$ change their U(1) charges under the discrete 
transformations.
The difference between the  U(1) gauge
interaction we are considering here and the electromagnetic one lies
in the fact that the states $\psi_u$ and $\psi_d$, having
one and the same electric charge $q$, are not distinguished by
the electromagnetic interaction.

The quantum counterparts of the classical constraints (\ref{62}) and
(\ref{63})
have the form
\begin{equation}
\hat{\tilde{\chi}}\Psi=0,
\quad
\hat{\tilde{\chi}}=\hat{\cal P}\gamma\otimes 1+
m\cdot 1\otimes\sigma_3,
\label{f1}
\end{equation}
\begin{equation}
\hat{\tilde{\phi}}\Psi=0,
\quad
2\hat{\tilde{\phi}}=
\hat{\cal P}^2+m^2
+\frac{1}{2}\epsilon_{\mu\nu\lambda}\gamma^\lambda\otimes
1\cdot \left(\frac{1}{2}{\cal QF}^{\mu\nu}\cdot 1\otimes
\sigma_3 +q F^{\mu\nu}\cdot 1\otimes 1\right),
\label{f2}
\end{equation}
where $\hat{\cal P}_\mu=p_\mu-{\cal Q}\hat{\cal N}{\cal
A}_\mu(x)-qA_\mu(x)$.  The operators $\hat{\tilde{\chi}}$ and
$\hat{\tilde{\phi}}$
satisfy the initial trivial algebra, i.e.
$[\hat{\tilde{\phi}},\hat{\tilde{\chi}}]=0$, and the equation
(\ref{chif}) is still valid
when the fermions are interacting with the chiral and
electromagnetic U(1) fields.

The field Lagrangian describing the interacting theory is
\begin{equation}
{\cal L}(x)=\overline{\Psi}(x)\hat{\tilde{\chi}}\Psi(x),
\label{lagf}
\end{equation}
where $\overline{\Psi}=\Psi^\dagger\cdot\gamma^0\otimes 1$,
and $p_\mu=-i\partial_\mu$.
The equations of motion derived from the Lagrangian
(\ref{lagf}) are given by eq. (\ref{f1}),
whereas the eq. (\ref{f2}), quadratic in $\partial_\mu$, 
comes as a consequence of equation (\ref{chif}).
The field Lagrangian (\ref{lagf}) has been used  
in some models of high-T${}_c$
superconductivity~\cite{8}.

\section{Concluding remarks}

In this paper we showed how the hidden U(1,1)
symmetry and the hidden $N=3$ supersymmetry
of the $P,T-$invariant quantum fermion model described 
by (\ref{i1}) could be understood in terms of the 3D
pseudoclassical model described by eqs. (\ref{2})--(\ref{6}).
In particular, our approach clarified the nature of the 
pseudounitary U(1,1) symmetry appearing in the quantum 
model since  --- at the classical level ---
the two pairs of oscillator-like odd variables
$\theta^\pm$ and $\xi^{(\pm)}$, from which the generators 
of the U(1,1) symmetry are constructed, have Poisson brackets 
of opposite sign.
{}Furthermore, the pseudoclassical model clearly shows that the generators
of the SU(1,1) symmetry and the odd generators of the $N=3$ supersymmetry
are intimately related; in fact, they only differ in the factor
$\xi^{(0)}$ which is an integral of motion.
For the quantum model, the generators of the SU(1,1)
symmetry and those of the $N=3$ supersymmetry
coincide up to a $c-$number factor since they are eigenvectors
of the operator $\hat{\xi}{}^{(0)}$ corresponding to the same 
eigenvalue.

The continuous symmetries of the pseudoclassical 
and $P,T-$invariant quantum fermion model are identical if 
one requires that the classical model has a maximal symmetry;
it is this requirement the one selecting the special value ($\nu=1$) 
of the parameter appearing in the pseudoclassical model.
It would be interesting 
to see how a special value of the 
parameter $\nu$ may be selected in the 
path-integral approach. Unfortunately, this is a difficult task \cite{pr} 
since the model has an even nilpotent constraint
for which a gauge fixing condition cannot be introduced.

Interesting features appear also when one supplement the
pseudoclassical model with a local U(1) gauge symmetry
which is the classical analog of the corresponding quantum  
U${}_c$(1) chiral gauge symmetry.
As it is shown in Appendix A,
when the parameter $\nu$  is not restricted by the condition
$\nu=1$, in the ungauged model
there is  a hidden symmetry  
realized in the sector of the $\theta^\pm$ variables,
which is responsible for a sort of 
quantization condition on the parameter $\nu$:
\[
\nu=\nu_{lk}=\frac{2l+1}{2k+1},\quad
l,k\in Z.
\]
This symmetry is effectively ``restored" in the gauge model
even when the maximal continuous symmetry of the  
free pseudoclassical model ($\nu=1$) is required.  
It seems to be interesting 
to look for the quantum analog of this hidden 
symmetry and reveal its
role in the context of the field 
model with the gauged U${}_c$(1) chiral symmetry.

The quantum generators of the N=3 supersymmetry
together with the energy-momentum vector and the total angular
momentum vector operators are the generators of a
superextension of the (2+1)-dimensional Poincar\'e group.
The complete set of possible eigenvalues  of the superspin operator,
which is one of the Casimir operators of this supergroup,
is given by the values $(-1/2,0,0,+1/2).$
The physical subspace of the quantum
$P,T-$invariant fermion model realizes an irreducible
representation of the Poincar\'e supergroup,   
labelled by the zero eigenvalue of the superspin.
In this representation the states of the $P,T-$invariant system
carry spin $-1/2$ and $+1/2$.

{}In the analysis of the 
U(1,1) symmetry given in section 3, either the variables 
$\theta_a$ or the variables $\xi^{(i)}=\xi^{\mu}e^{(i)}_\mu$ are not
U(1,1) covariant quantities; only their specific linear
combinations $\Sigma^{i}_a$, defined in eq. (\ref{spinor}), are covariant
under the action of U(1,1).
One may then expect that the model 
(\ref{2})--(\ref{6}), written in terms of the latter set of variables,
is manifestly 
U(1,1) covariant and shows explicitly the 
hidden N=3 supersymmetry.
However, since $\{x_\mu,\Sigma^{i}_a\}\neq 0$,
the reformulation of the system
in a manifestly U(1,1) covariant way is a nontrivial task
within the Lagrangian formalism.

In conclusion, the
simple pseudoclassical model considered in this paper
reveals a nontrivial `superposition' of
discrete and continuous, global and local, ordinary and
graded symmetries, and allows for 
a natural way 
to construct a field Lagrangian with a U(1) gauge field coupled
to the chiral current. One of our motivation to investigate the properties of 
this particular pseudoclassical model lies in the fact 
that the planar SPM \cite{3} does not naturally reproduce 
at the classical  
level the symmetries of the $P,T-$invariant 3D quantum fermion model
of ref. \cite{7}. Moreover,
in the SPM there is no natural analog of the U(1) 
chiral symmetry we explicitly constructed, since this symmetry
acts only in the subspace of the $\theta^{\pm}$ variables.
Due to this, a gauge generalization based on the SPM is not 
immediate. Thus,
the pseudoclassical model of ref.
\cite{9} provides the simplest classical example of a model 
exhibiting the symmetries of the $P,T-$invariant 
planar quantum free fermion model.
\vskip1.0cm 

{\bf Acknowledgements}

One of us (M.P.) is grateful to J.L. Cort\'es for discussions.
The research of M.P. was supported by MEC-DGICYT (Spain).
He also thanks INFN for financial support and the University
of Perugia for hospitality.

\appendix

\section{Hidden symmetry}

There is a `hidden' symmetry in the free model
(\ref{2}), induced by transformations in the
subspace of the variables $\theta^\pm$. It is a product of one
of the discrete transformations defined by eq. (\ref{19}), say
the $P$ inversion, and the special U(1) transformation
\begin{equation}
U_{s}:\, \theta^\pm \rightarrow
{\theta^\pm}'=e^{\pm i{\gamma}_s(\tau)}\theta^\pm,
\label{81}
\end{equation}
with $\gamma_s(\tau)$ given by: 
\begin{equation}
{\gamma}_s(\tau)=-2\nu\Omega(\tau),\quad
\Omega(\tau)=2m\int_{\tau_i}^{\tau}v(\tau')d\tau'.
\label{82}
\end{equation}
In (\ref{82}) it is assumed that $\nu\neq 0$.  
Under the transformation $U_s P_\theta$,
 --- $P_\theta$ being a restriction of the $P$ inversion to the
sector of the variables $\theta^\pm$,
i.e. $P_\theta :\theta_a\rightarrow \theta'_a=(\theta_1,-\theta_2)$ ---
the Lagrangian is invariant: $U_s P_\theta: L\rightarrow L$.
The boundary term $B$ is
invariant under this transformation $U_s P_\theta$
iff the function ${\gamma}_s(\tau)$ is such that
\begin{equation}
\Delta\gamma_s=\gamma_s(\tau_f)-\gamma_s(\tau_i)=\gamma(\tau_f)=
2\pi n,
\quad
n\in Z,
\label{83}
\end{equation}
which amounts to 
\begin{equation}
\nu \Omega(\tau_f)=\pi n, \quad
n\in Z.
\label{84}
\end{equation}
This condition is not affected 
by the local transformations
(\ref{14}) and (\ref{15}) due to the boundary conditions (\ref{16})
imposed on the parameters $\alpha$ and $\beta$.

In accordance with eqs. (\ref{26}), (\ref{39}), one gets
\begin{equation}
\xi^{(\pm)}(\tau_f)=\xi^{(\pm)}(\tau_i)\cdot e^{\pm
i\Omega(\tau_f)},\qquad
\theta^{\pm}(\tau_f)=\theta^{\pm}(\tau_i)\cdot
e^{\pm i\nu\Omega(\tau_f)}.
\label{85}
\end{equation}
The boundary conditions (\ref{7}) require that 
the quantities
\[
\theta^\pm_{1/2}=\theta^\pm(\tau_i)+\theta^\pm(\tau_f),\qquad
\xi^{(\pm)}_{1/2}=\xi^{(\pm)}(\tau_i)+\xi^{(\pm)}(\tau_f)
\]
must be fixed. In terms of the above variables, one has that
\begin{equation}
\xi^{(\pm)}(\tau_i)=\left(1+e^{\pm
i\Omega(\tau_f)}\right)^{-1}\cdot \xi^{(\pm)}_{1/2},\quad
\xi^{(\pm)}(\tau_f)=\left(1+e^{\mp
i\Omega(\tau_f)}\right)^{-1}\cdot \xi^{(\pm)}_{1/2},
\label{86}
\end{equation}
and similar equations for $\theta^\pm$ with the corresponding
change of $\Omega(\tau_f)$ in $\nu\Omega(\tau_f)$.
When $\xi^{(\pm)}_{1/2}\neq 0$ or (and)
$\theta^\pm_{1/2}\neq 0$, and
one (or both) of the equations
\begin{equation}
\Omega(\tau_f)=\pi(2k+1),\qquad
\nu\Omega(\tau_f)=\pi(2l+1),\quad
k,l\in Z,
\label{87}
\end{equation}
hold, (\ref{86}) cannot be staisfied since the denominators 
vanish.
Thus, either one prohibits
$\Omega(\tau_f)$ to obey
eqs.~(\ref{87}), or one requires that
\begin{equation}
\theta^\pm_{1/2}=0,\quad
\xi^{(\pm)}_{1/2}=0.
\label{88}
\end{equation}
If (\ref{88}) is satisfied,
the conditions (\ref{87}) must be also satisfied. 
We choose the second
possibility, i.e. we assume 
antiperiodic boundary (in evolution parameter)
conditions for the Grassmann `oscillating' variables
$\theta^\pm(\tau)$ and $\xi^{(\pm)}(\tau)$.
The conditions (\ref{87})
lead to a sort of quantization condition $\nu=\nu_{lk}$,
\begin{equation}
\nu_{lk}=\frac{2l+1}{2k+1},\quad
k,l\in Z,
\label{89}
\end{equation}
which includes the special case $\nu^2=1$.
For the set of values of the parameter $\nu$ given by eq.
(\ref{89}), the boundary conditions (\ref{84}) are satisfied,
and, hence, the hidden symmetry described above is realized.

For $\nu=1$, the model described by the
Lagrangian $\tilde{L}=L+L_{int}+L^{em}_{int}$,
where $L$, $L_{int}$ and $L_{int}^{em}$ are given by
eqs. (\ref{2}), (\ref{li}) and (\ref{liem}), 
is invariant under the hidden
symmetry $U_s P_{\theta}$, if one supplements the transformation 
properties of the pseudoscalar variables with the transformation
law $U_s P_{\theta}: ({\cal A}_{\mu}, A_\mu )\rightarrow
(-{\cal A}_{\mu},A_\mu)$ for the U(1)
gauge fields.

\section{Standard $3D$ pseudoclassical model}

Here we shall show that, though the quantization of the 
standard pseudoclassical model for 
the massive relativistic spin particle \cite{3}
also leads to the $P,T-$invariant 3D fermion quantum model,
nevertheless, the SPM does not reproduce either the  
U(1,1) symmetry or the $N=3$ supersymmetry
at the classical level.  

The Lagrangian of the model \cite{3} is similar in form
to the one of the model we have considered in this paper,
namely:
\[
L=\frac{1}{2e}(\dot{x}_\mu-i\lambda\xi_\mu)^2-\frac{e}{2}m^2
-im\lambda\xi_*-\frac{i}{2}\xi_\mu\dot{\xi}{}^{\mu}
-\frac{i}{2}\xi_*\dot{\xi}_*.
\]
There is an odd Lagrange multiplier $\lambda$
instead of the even multiplier $v$, and one odd (pseudo)scalar
variable $\xi_*$ instead of the pair of variables  
$\theta_a$, $a=1,2$. 
The total Hamiltonian of the system is given by:
\[
H=e\phi+i\lambda\chi+up_e+\rho\pi_\lambda.
\]
It is a linear combination of the primary constraints 
$p_e\approx0$ and $\pi_\lambda\approx0$, $\pi_\lambda$
being the canonical momentum conjugate to $\lambda$,
of the constraint $\phi$ given by eq. (\ref{22}) and of
\begin{equation}
\chi=p\xi+m\xi_*\approx0.
\label{a1}
\end{equation}
The brackets of $\xi_\mu$ are given by the same eq. (\ref{20}),
whereas
\[
\{\xi_*,\xi_*\}=i.
\] 
The equations of motion for the odd variables are 
$\dot{\xi}_\mu=\lambda p_\mu$ and $\dot{\xi}_*=m\lambda$.
The quantities
$
\xi^{(i)},$ $i=1,2,$ and $\xi^{(0)}\xi_*$ are integrals of motion.
The last even integral is weakly equal to zero if one takes into account
the constraint (\ref{a1}), but it is a nontrivial operator
at the quantum level.

Let us briefly consider the quantum theory of this model,
and then return to the classical model in order to construct the 
classical counterparts of the quantum symmetry generators.
The quantum operators corresponding to the variables
$\xi_\mu$ and $\xi_*$ may be realized as
\begin{equation}
\hat{\xi}_\mu=\frac{1}{\sqrt{2}}\gamma_\mu\otimes\sigma_1,
\quad
\hat{\xi}_*=-\frac{i}{\sqrt{2}}1\otimes \sigma_2.
\label{a3}
\end{equation}
The internal scalar product is defined as
$(\Psi_1,\Psi_2)=\bar{\Psi}_1\Psi_2$,
with $\bar{\Psi}=\sqrt{2}\Psi^{+}\hat{\xi}{}^{0}$.
With this indefinite scalar product
all the operators (\ref{a3}) are hermitian.
The quantum counterpart of the constraint (\ref{a1})
is 
\begin{equation}
\frac{1}{\sqrt{2}}1\otimes\sigma_1(p\gamma\otimes 1+m 1\otimes\sigma_3)\Psi=0,
\label{a4}
\end{equation}
which is equivalent to eq. (\ref{71}).
If one takes 
${\cal L}=\sqrt{2}\bar{\Psi}\hat{\chi}\Psi$ as the field
Lagrangian,  
one finds that it coincides with the Lagrangian (\ref{lagf})
(with $A_\mu={\cal A}_\mu=0$).
Hence, the quantization of the SPM reproduces the 
$P,T-$invariant 3D fermion system.  

The classical quantities corresponding to the generators (\ref{76})
are given by :
\begin{equation}
R_0=\frac{i}{2}\xi^{(1)}\xi^{(2)}\cdot G,
\quad
R_1=\frac{1}{2\sqrt{2}}\xi^{(2)}\cdot G,
\quad
R_2=-\frac{1}{2\sqrt{2}}\xi^{(1)}\cdot G,
\label{a5}
\end{equation}
where $G=1-2\xi^{(0)}\xi_*$.
First of all, one sees that these quantities
have different Grassmann parity:
$R_0$ is even, whereas $R_i$, $i=1,2$, are odd;
in addition,  one can check that they do not form an algebra
with respect to the brackets. 
The classical counterpart of the operator $\gamma^{(0)}\otimes\sigma_3$
is, up to a c-number factor, the even quantity $\xi^{(0)}\xi_*$;
therefore, one has no possibility to `change' the Grassmann
parity of the integrals of motion as it was possible 
for the model (\ref{2}). Due to this, in this model one cannot reproduce the 
$N=3$ supersymmetry at the classical level.

One may start not from the quantum constraint
given by eq. (\ref{71}) and equivalent to equation (\ref{a4}), 
but from the quantum constraint (\ref{a4}) itself, which is equal
to the constraint (\ref{71}) multiplied by $1\otimes \sigma_1$.
One can then take the corresponding operator
$\hat{\tilde{\cal X}}=\gamma^{(0)}\otimes\sigma_1-i 1\otimes\sigma_2$
as the generator of the U(1) symmetry.
In this case one has the following set of generators of the SU(1,1)
symmetry, commuting with the U(1) generator:
\begin{equation}
\Gamma^{(0)}=-\frac{1}{2}\gamma^{(0)}\otimes 1,\quad
\Gamma^{(i)}=-\frac{1}{2}\gamma^{(i)}\otimes\sigma_2.
\label{a6}
\end{equation}
From these one constructs the Lorentz vector $\Gamma_\mu =e^{(\alpha)}_\mu
\Gamma_{(\alpha)}$ satisfying the equation
\[
\Gamma_{\mu}\Gamma_{\nu}=-\frac{1}{4}\eta_{\mu\nu}
-\frac{i}{2}\epsilon_{\mu\nu\lambda}\Gamma^{\lambda}.
\]
The anticommutator of these operators is a constant;
the nontrivial supersymmetry of the generators is lost.
Due to this fact,
instead of (\ref{*s}), in the SPM one has 
$\hat{\cal S}=\hat{\cal J}{}^{(0)}=0$, 
$\hat{\cal J}_\mu=
\hat{J}_\mu-\Gamma_\mu$. 
Again, the classical analogs of the SU(1,1)
generators (\ref{a6}), given by 
$\Gamma^{(0)}=-i\xi^{(1)}\xi^{(2)}$ 
and $\Gamma^{(i)}=-\sqrt{2}\epsilon^{ij}\xi^{(0)}\xi^{(j)}\xi_*$,
do not reproduce the SU(1,1) algebra at the classical level.

In the SPM 
there is no natural analog of the U(1) symmetry
acting in the subspace of the variables
$\theta^\pm$.
Furthermore, using the SPM, it is cumbersome to
provide a natural construction of the U(1) gauge theory.
Thus, the SPM does not reproduce the symmetries
of the quantum $P,T-$invariant 3D fermion theory at the classical level.

\end{document}